\documentclass[final,3p,times,a4paper]{elsarticle}

\PassOptionsToPackage{margin=1in}{geometry}
\usepackage{amsmath,amssymb} 
\usepackage{graphicx} 
\usepackage{xspace} 
\usepackage{hyperref}
\hypersetup{
colorlinks = true,
linkcolor = blue,
anchorcolor = blue,
citecolor = blue,
filecolor = blue,
urlcolor = blue
}
 
\newcommand{\sNN}{\sqrt{s_{\rm NN}}\xspace} 
\newcommand{\muB}{\mu_B} 
\newcommand{\muQ}{\mu_Q} 
\newcommand{\muS}{\mu_S} 
\newcommand{\pt}{p_T} 
\newcommand{\RAA}{R_{\rm AA}}

\journal{Journal of Subatomic Particles and Cosmology}

\begin{document}

\begin{frontmatter}

\title{Multistage dynamical modeling of heavy-ion collisions}

\author[first,second]{Lipei Du}
\ead{ldu2@lbl.gov}
\fntext[label1]{Invited plenary talk at the 22nd International Conference on Strangeness in Quark Matter, Los Angeles, CA, USA, March 22–27, 2026.}
\affiliation[first]{organization={Department of Physics, University of California},
            city={Berkeley},
            postcode={94270}, 
            state={CA},
            country={USA}}
\affiliation[second]{organization={Nuclear Science Division, Lawrence Berkeley National Laboratory},
            city={Berkeley},
            postcode={94270}, 
            state={CA},
            country={USA}}
\date{\today}

\begin{abstract}
Relativistic heavy-ion collisions create deconfined QCD matter whose properties must be inferred from final-state observables through dynamical modeling.  This contribution discusses recent progress and open issues in multistage simulations, with emphasis on the connection between bulk evolution, conserved charges, strangeness, and heavy flavor.  At RHIC Beam Energy Scan energies, the breaking of longitudinal boost invariance makes charge stopping and rapidity-dependent observables essential for constraining the finite-density medium.  Strange hadrons are sensitive to the local chemical environment and conserved-charge correlations, while heavy flavor probes microscopic transport and hadronization.  Combining these observables within multi-sector inference frameworks provides a path toward more robust constraints on the equation of state and transport properties of QCD matter.
\end{abstract}


\end{frontmatter}

\section{Introduction}

Relativistic heavy-ion collisions provide a laboratory for creating and studying deconfined QCD matter under extreme conditions~\cite{Arslandok:2023utm,Sorensen:2023zkk}.  The central theoretical task is to connect the final-state particles measured in experiment to the space-time evolution and microscopic properties of the quark--gluon plasma (QGP).  This connection requires solving two related problems.  The forward problem asks how an evolving QCD medium produces and modifies hadrons, electromagnetic radiation, jets, and heavy flavor.  The inverse problem asks how these observables can be used to constrain the equation of state (EoS), transport coefficients, initial conditions, and microscopic interaction mechanisms of the medium~\cite{Paquet:2023rfd}.

Dynamical modeling is the framework in which these two problems meet.  A realistic description of a collision involves a sequence of stages: fluctuating initial energy and charge deposition, pre-equilibrium evolution, relativistic viscous hydrodynamics, particlization, and hadronic transport.  Each stage affects different observables with different sensitivities.  Therefore, the interpretation of strangeness and heavy flavor should not be separated from the bulk evolution.  Strange hadrons carry information about the chemical environment, conserved-charge correlations, and hadronization, while heavy quarks are produced at the earliest times and probe the medium through diffusion, energy loss, and recombination~\cite{Apolinario:2022vzg,JETSCAPE:2022hcb}.

This contribution focuses on recent progress and open issues in dynamical model simulations of heavy-ion collisions, with emphasis on the connection between bulk evolution and flavor-sensitive probes.  At RHIC Beam Energy Scan (BES) energies, this connection requires three-dimensional finite-density dynamics, conserved-charge transport, and inference frameworks that combine complementary observables within a common medium description~\cite{Du:2024wjm}.

\section{Multistage dynamical modeling}

\begin{figure}
    \centering
    \includegraphics[width=\linewidth]{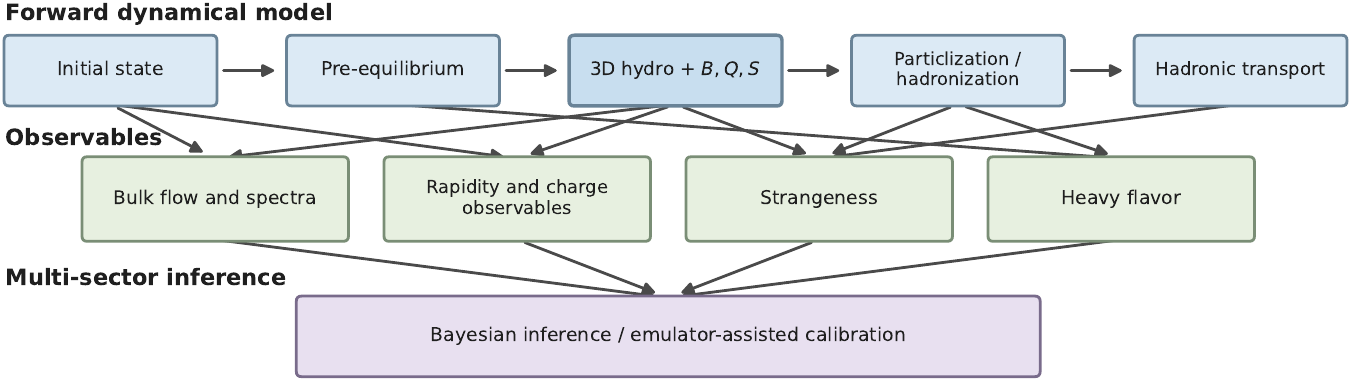}
    \caption{
    Schematic view of the forward and inverse problems in heavy-ion collisions. A multistage dynamical model evolves the system from the initial state to hadronic transport, while bulk, rapidity-dependent, strange, and heavy-flavor observables provide complementary constraints through multi-sector inference.
    }
    \label{fig:multi_sector}
\end{figure}

Figure~\ref{fig:multi_sector} summarizes the modeling logic emphasized in this contribution.  The forward description begins with fluctuating event-by-event distributions of energy, momentum, and conserved charges.  Energy-density fluctuations seed anisotropic flow and determine the background through which hard probes and heavy flavor propagate~\cite{Arslandok:2023utm,Du:2024wjm}.  Local fluctuations of baryon number, electric charge, and strangeness can provide an initial chemical environment that is later transported by hydrodynamics and sampled by strange hadrons~\cite{Gardim:2024nyz}.

After the initial impact, the system evolves through a pre-equilibrium stage before viscous hydrodynamics becomes applicable.  Although short-lived, this stage can leave visible imprints on early probes, especially heavy quarks and electromagnetic radiation.  Heavy quarks are created at times of order $\tau\simeq 0$ fm/$c$ through hard processes and therefore begin to sample the fluctuating medium before the system is fully hydrodynamized~\cite{Apolinario:2022vzg,JETSCAPE:2022hcb}.

A related open question is whether the early QGP is chemically equilibrated.  If the plasma is initially gluon dominated, the quark fugacity evolves during the subsequent expansion.  Such a time-dependent quark composition can modify entropy production and influence hadron yields and flows.  This provides a direct connection between early-time microscopic chemistry and final-state strangeness production~\cite{Gordeev:2025vog}.

The hydrodynamic stage converts pressure gradients into collective flow and, at finite density, transports conserved charges.  Near particlization and in the hadronic phase, the evolving medium is mapped into the final flavor composition and momentum distributions~\cite{Arslandok:2023utm,Apolinario:2022vzg}.

Viewed from the probe perspective, this multistage evolution implies that different observables carry information over different time windows, as illustrated schematically in Fig.~\ref{fig:multi_sector}.  Heavy quarks sample nearly the entire evolution, strange hadrons are sensitive to both chemical production and conserved-charge transport, and final hadron distributions encode the late-time particlization and hadronic phases.  This time ordering helps flavor-sensitive observables distinguish contributions from different stages of the collision.

\section{Three-dimensional dynamics and charge stopping}

The limitations of boost-invariant modeling become most visible when one follows the rapidity dependence of particle production.  At LHC energies and near midrapidity, the QGP is approximately boost invariant, and many bulk observables can be described using $(2+1)$-dimensional hydrodynamic simulations.  This approximation becomes inadequate at BES energies~\cite{Du:2024wjm}.  Charged-particle, identified-particle, and net-baryon rapidity distributions show strong longitudinal structure, implying that the full three-dimensional evolution of the fireball must be modeled~\cite{Du:2023efk}.

Charge stopping, especially baryon stopping, is a central ingredient in this regime because it determines the net-baryon distribution and therefore the finite-$\muB$ region explored by the collision.  More generally, the incoming nuclei carry conserved charges into the collision, and their longitudinal deceleration determines how net charge densities are deposited in rapidity.  These initial rapidity profiles are subsequently modified by hydrodynamic expansion, coupled charge diffusion, particlization, and hadronic transport~\cite{Denicol2018}.  Final-state rapidity-dependent observables therefore retain information about both the early stopping mechanism and the later transport of conserved charges.

It is useful to view the longitudinal dynamics as a multistage mapping.  The initial stage fixes the rapidity profiles of entropy and conserved charges.  The hydrodynamic stage converts longitudinal pressure gradients into longitudinal flow, while gradients of the charge densities and chemical potentials drive diffusion.  Near freeze-out, longitudinal flow further reshapes the final rapidity distributions.  Starting from the same underlying density profile, stronger longitudinal flow broadens the distribution, with heavier particles experiencing a larger kinematic stretching~\cite{Du:2023gnv}.  A realistic description of BES observables therefore requires consistency among initial charge deposition, hydrodynamic transport, longitudinal flow, and particlization~\cite{Du:2024wjm}.

Rapidity-dependent measurements provide a complementary way to explore the QCD phase diagram.  In the usual beam-energy scan, changing $\sNN$ changes the average trajectory of the fireball in the $(T,\muB)$ plane.  In a rapidity scan, different rapidity regions within a single collision sample different local thermodynamic conditions~\cite{Du:2023gnv}.  This should not be interpreted as a sharp point-by-point mapping from rapidity to thermodynamic parameters: thermal smearing and the finite width of the freeze-out hypersurface imply that extracted thermal properties are averaged over a rapidity window.  Nevertheless, rapidity dependence provides an important handle on the longitudinal variation of temperature and chemical potentials.

Directed flow provides a particularly sensitive probe of three-dimensional dynamics.  The rapidity-odd coefficient $v_1(y)$ is generated early and responds to the tilted or shifted geometry of the expanding fireball, the distribution of baryon density in the reaction plane, and the longitudinal pressure gradients.  The midrapidity slope of baryon directed flow is therefore a diagnostic of the interplay between baryon stopping and collective expansion.  Nontrivial structures in baryon $v_1(y)$ can arise from this three-dimensional dynamics, making realistic 3D baselines essential before attributing directed-flow structures to the EoS or phase structure~\cite{Du:2022yok}.

\section{Conserved charges and finite-density hydrodynamics}

At finite baryon density, hydrodynamics should evolve not only energy and momentum but also the conserved currents associated with baryon number $B$, electric charge $Q$, and strangeness $S$.  These currents can be decomposed as
\begin{equation}
  \partial_\mu N_q^\mu = 0, \qquad
  N_q^\mu = n_q u^\mu + V_q^\mu, \qquad q \in \{B,Q,S\},
\end{equation}
where $n_q$ is the local charge density and $V_q^\mu$ is the diffusion current~\cite{Denicol2018,Du:2019obx}.  Schematically, the diffusion currents satisfy constitutive equations of the form
\begin{equation}
  \sum_{q'} \tau_{qq'} V_{q'}^{\langle\mu\rangle}
  + V_q^\mu
  =
  \sum_{q'} \kappa_{qq'} \nabla^\mu \left( \frac{\mu_{q'}}{T} \right)
  + \cdots ,
\end{equation}
where $\kappa_{qq'}$ is the conductivity matrix~\cite{Greif:2017byw}.  The off-diagonal components mean that a gradient in one thermal potential, such as $\muB/T$, can induce diffusion of other charges.  This coupled structure is essential for a consistent treatment of charge transport in QCD matter.

The finite-density EoS, $p=p(T,\muB,\muQ,\muS)$, links the local charge densities to the corresponding chemical potentials and therefore also enters the diffusion dynamics~\cite{Monnai2019,Abuali:2025tbd}.  At small chemical potentials, lattice QCD constrains this EoS through Taylor expansions of the pressure.  The same conserved-charge susceptibilities that enter this expansion also encode correlations among $B$, $Q$, and $S$.  At BES energies, one must implement this information in a form suitable for dynamical simulations, including the treatment of local strangeness neutrality and the relation between electric charge and baryon number~\cite{Du:2023efk,Monnai2019}.

Although the total strangeness of the collision system vanishes, the local strangeness density and the associated chemical potential need not be trivial during the evolution.  The initial BQS fluctuations discussed above, together with the coupled BQS EoS, can generate local correlations among baryon number, electric charge, and strangeness.  Multi-charge hydrodynamics then transports these correlations and can translate them into measurable differences among identified hadrons~\cite{Gardim:2024nyz}.  In this sense, strange hadrons carry finite-density information about the local chemical environment, not only information about final-state abundances.

Particle ratios provide a simple illustration.  Schematically, in a thermal picture,
\begin{equation}
  K^-/K^+ \sim \exp[-2(\muQ+\muS)/T], \qquad
  \bar p/p \sim \exp[-2\muB/T].
\end{equation}
Thus kaons and protons probe different combinations of chemical potentials.  Their rapidity dependence constrains the longitudinal profiles of $\muB$, $\muQ$, and $\muS$, and therefore the underlying charge stopping and transport~\cite{Du:2023efk,Monnai2019}. 
More generally, ratios constructed from conserved-charge carriers can provide sensitivity to the relative stopping and transport of different conserved charges, including baryon number, electric charge, and strangeness~\cite{Ross:2025qxr}.  Such observables are useful because they probe the longitudinal charge structure more directly than inclusive charged-particle production.

Complementary information comes from identified-particle flow, which provides a more differential probe of the finite-density EoS.  Kaons and lambdas respond to different chemical-potential combinations, such as $\muQ+\muS$ for charged kaons and $\muB-\muS$ for lambdas.  Their directed flow can therefore be sensitive to the treatment of strangeness neutrality, charge conservation, and the finite-density EoS~\cite{Du:2022yok}.  This species dependence complements the use of baryon directed flow as a probe of three-dimensional geometry and stopping dynamics.

\section{Strangeness and heavy flavor as dynamical probes}

Strangeness and heavy flavor extend the finite-density discussion to complementary probes of hadronization and microscopic transport.  While strange hadrons connect naturally to the conserved-charge dynamics discussed above, heavy flavor provides an independent handle on how energetic and massive probes interact with the evolving medium.

Beyond the finite-density sensitivity discussed above, strangeness also probes the later stages of the collision.  Canonical and local charge conservation effects~\cite{Ciacco:2026odz} can become important in small systems, local domains, or for rare species, while strange resonances provide sensitivity to the hadronic phase through rescattering and regeneration.  Strangeness therefore connects conserved-charge transport to the hadronization and hadronic stages of the evolution.

For heavy flavor, the main dynamical information is encoded in how charm and bottom quarks interact with the evolving QGP~\cite{Apolinario:2022vzg}.  The nuclear modification factor $\RAA$ constrains energy loss, while elliptic flow $v_2$ constrains how efficiently heavy quarks participate in the collective expansion.  Heavy-flavor hadronization adds another layer: recombination with thermal light or strange quarks at low and intermediate $\pt$ can modify charm-hadron chemistry and make baryon-to-meson and strange-to-nonstrange heavy-flavor ratios sensitive to the partonic environment near hadronization~\cite{Apolinario:2022vzg,JETSCAPE:2022hcb}.

A useful future direction is to develop observables that correlate sectors often analyzed separately.  The fluctuation observable $v_0(\pt)$ is one example.  For identified light and strange hadrons, $v_0(\pt)$ characterizes how spectra fluctuate in correlation with the event-wise mean transverse momentum~\cite{Du:2025dpu} and can reveal the response of spectral shapes to radial-flow and hard-component fluctuations.  For heavy flavor, analogous observables may be sensitive to transport coefficients and hadronization~\cite{Sambataro:2025pop}.  Such soft--hard and flavor-dependent correlations are valuable because they test how bulk fluctuations are transmitted to microscopic probes, thereby helping to break degeneracies that remain in single-sector analyses~\cite{Du:2025hrz}.

\section{Toward quantitative multi-sector inference}

The preceding sections highlight a common challenge: different observables constrain different aspects of the same evolving medium.  The inverse problem is therefore intrinsically many-to-many: several probe sectors share the same bulk evolution, while each sector can introduce its own transport or hadronization parameters.  Bayesian inference provides a natural framework for propagating uncertainties and constraining these parameters with multiple observables~\cite{Paquet:2023rfd}.

A central issue, illustrated in Fig.~\ref{fig:multi_sector}, is how to combine different sectors.  One possible strategy is sequential inference~\cite{Roch:2025jpu,Du:2026dvl}: first constrain the bulk medium using soft observables, then propagate the posterior ensemble of medium evolutions into calculations of heavy flavor, jets, electromagnetic probes, or conserved-charge observables.  This approach is computationally efficient and makes the uncertainty from the bulk medium explicit.  However, it assumes that the additional sectors do not significantly change the preferred medium parameters.  A fully joint inference, in which soft and non-soft observables constrain the medium and probe-specific transport coefficients simultaneously, is conceptually more complete but computationally more demanding~\cite{Bernhard:2019bmu,JETSCAPE:2020shq,JETSCAPE:2020mzn,Du:2025rgq}.

This distinction is not only statistical.  If hard, heavy-flavor, electromagnetic, or conserved-charge observables prefer a different region of medium-parameter space than the soft sector alone, this may point to missing correlations or missing physics in the model.  Conversely, if a common medium evolution can describe bulk flow, strangeness, heavy flavor, and rapidity-dependent observables simultaneously, the extracted QGP properties become more robust.  Multi-sector consistency is therefore a physics constraint, not merely a fitting strategy.

New observables should be evaluated by the degeneracies they help break.  Rapidity-dependent particle ratios can separate the effects of charge stopping and local chemical potentials.  Identified-particle directed flow can help disentangle longitudinal geometry from finite-density equation-of-state effects.  Soft--hard fluctuation observables such as $v_0(\pt)$ can test how bulk fluctuations correlate with microscopic transport and hadronization.  Such observables are most powerful when embedded in a common inference framework.

Such analyses are computationally demanding because they require repeated event-by-event evaluations of a multistage model in a high-dimensional parameter space.  Fast emulators and surrogate models are therefore essential, but they must be validated against the underlying dynamical simulations and used with controlled uncertainties~\cite{Paquet:2023rfd}.

\section{Summary and outlook}

Dynamical simulations provide the bridge between QCD matter properties and heavy-ion collision observables.  At BES energies, this bridge must be fully three-dimensional and finite-density: charge stopping, longitudinal expansion, and conserved-charge transport are essential ingredients.  In this regime, strangeness carries information about the local chemical environment, while heavy flavor probes microscopic transport and hadronization.

The main challenge is to turn this multistage, multi-probe picture into quantitative constraints.  This requires realistic initial conditions for entropy and conserved charges, finite-density equations of state and transport coefficients, consistent flavor hadronization, and inference frameworks that preserve correlations across sectors.  Such multi-sector analyses offer a path toward a more precise determination of QCD matter properties.

~\\
\noindent {\bf Acknowledgments.} The author thanks the organizers of Strangeness in Quark Matter 2026 for the invitation to present this overview, and acknowledges useful discussions with collaborators and participants of the conference.  This work was supported in part by the U.S. Department of Energy, Office of Science, Office of Nuclear Physics under Grant No. DE-AC02-05CH11231.

\bibliographystyle{elsarticle-num}
\bibliography{refs}

\end{document}